\documentclass[conference]{IEEEtran}

\usepackage{graphicx}
\usepackage[cmex10]{amsmath}
\usepackage{amssymb}
\usepackage{amsfonts}
\usepackage{amsthm}
\usepackage{amsmath,bm}
\usepackage{mathptmx} % pretty math
\usepackage[]{cite}
\usepackage{subfigure}
\usepackage{epstopdf}
\graphicspath{{figs/},{plots/}}
\usepackage{color}
\newcommand{\esn}[2] {ESN$_{\text{#1},\text{#2}}$}
\begin{document}
\title{A Model for Variation- and Fault-Tolerant Digital Logic using Self-Assembled Nanowire Architectures}

\author{\IEEEauthorblockN{Alireza~Goudarzi,~Matthew~R.~Lakin,~and~Darko~Stefanovic}
\IEEEauthorblockA{Department
of Computer Science\\
University of New Mexico\\
Albuquerque, NM 87131, USA\\
 Email: alirezag@cs.unm.edu}
\and
\IEEEauthorblockN{Christof Teuscher}
\IEEEauthorblockA{Department of Electrical\\ and Computer Engineering\\
Portland State University\\
Portland, OR 97201, USA}
}

\maketitle

\begin{abstract}
Reconfiguration has been used for both defect- and fault-tolerant nanoscale architectures with regular structure. Recent advances in self-assembled  nanowires have opened doors to a new class of electronic devices with irregular structure. For such devices, reservoir computing has been shown to be a viable approach to implement computation. This approach exploits the dynamical properties of a system rather than specifics of its structure. Here, we extend a  model of reservoir computing, called the echo state network, to reflect more realistic aspects of self-assembled nanowire networks. As a proof of concept, we use echo state networks to implement  basic building blocks of digital computing: AND, OR, and XOR gates, and 2-bit adder and multiplier circuits. We show that the system can operate perfectly in the presence of variations five orders of magnitude higher than ITRS's 2005 target, $\bm{6\%}$, and achieves success rates $\bm{6}$ times higher than related approaches at half the cost. We also describe an adaptive algorithm that can detect faults in the system and reconfigure it to resume perfect operational condition.
\end{abstract}

\IEEEpeerreviewmaketitle

\section{Introduction}
%%% no \IEEEPARstart

In view of approaching physical limits of silicon-based electronics, advances in materials science and nanotechnology suggest that unconventional computer architectures could be viable alternatives to current architectures. Some  proposed alternative architectures are based  on molecular switches and memristive crossbars \cite{Snider:2005qa} that possess  highly regular structures. Another emerging approach is the self-assembly of nanowires and memristive networks  \cite{ADMA:ADMA201103053}, which results in irregular structures. Here, we use a modification to a recent paradigm called {\em reservoir computing} (RC) to use the dynamics of current in a disordered system to compute digital functions. In contrast to standard approaches, our version of RC uses highly contractive dynamics to implement combinational logic perfectly, even in the presence of faults and temporal variations.

Major obstacles to using such architectures are design variations, defects, faults, and a susceptibility to environmental factors, such as thermal noise and radiation \cite{itrs2011}. Currently, common solutions  assume knowledge of the underlying architecture and rely on   reconfiguration and redundancy to achieve programming and fault tolerance\cite{1230995,6144380}. There have been two recent proposals on how to use such devices without knowledge of the underlying system\cite{Lawson:2006-04-01T00:00:00:1546-1955:272,6144633}. Both proposals are based on a model called a {\em randomly assembled computer} (RAC). In this model, a network of randomly and sparsely interconnected nodes with diode-like behavior is assumed. Three types of terminals connect the networks to the outside world: inputs, outputs, and controls. Internally, the terminals are connected to randomly chosen nodes. Using adaptive strategies one  finds  suitable control signals such that for a given input signal, the system produces a desired output. Computation in this approach is sensitive to the initial state of the network and therefore it is not readily applicable to time-dependent input signals. In addition, the control signals are calculated for a desired target, making the approach suitable for a single-purpose device only.

%%%%%%%%
%%%%%%%% Reservoir sample
%\begin{figure*}[th]
%\centering
%\includegraphics[]{Fig_rcdyn}
%\caption{Computation in a reservoir computer. The reservoir is made up of a dynamical neural network with randomly assigned weights. The states of the nodes are represented by ${\bf x}(t)$. The input signal ${\bf u}(t)$ is fed
%into every reservoir node $i$ with a corresponding weight $w^{in}_i$
%denoted by the weight column vector ${\bf W}^{in}=[w^{in}_i]$. Reservoir nodes are
%themselves coupled with one another using the weight matrix ${\bf W}^{res}
%=[w^{res}_{ij}]$, where $w^{res}_{ij}$ is the weight of the connection from node $j$ to node $i$.}
%\label{fig:fig_0}
%\end{figure*}

A new paradigm  that is gaining popularity in the unconventional computer architecture community is reservoir computing \cite{springerlink:10.1007}. In this approach, computation takes place in the transient dynamics of an excitable system, called a {\em reservoir}. The reservoir is perturbed by an external signal; the signal leaves an imprint in the dynamics of the system that can be translated into a desired output. The translation is performed by a linear readout layer, which can be trained efficiently using closed-form regression. The computational power of the reservoir is attributed to a short-term memory
created by the reservoir and the ability to preserve the
temporal information
from distinct signals over time\cite{springerlink:10.1007}. This property is known to be optimal in the critical dynamical regime of the reservoir---a regime
in which perturbations to the system's trajectory in its phase space neither spread nor
die out. It
has been suggested that the reservoir dynamics acts like a spatiotemporal
kernel, projecting the input signal onto a high-dimensional feature space\protect
\cite{Hermans:2011fk}.

In simulation, reservoir computing has been shown to be a viable approach for analog computation using self-assembled memristive networks \cite{6464167,6623028}, and recently, a physical implementation using self-assembled nanowires has also been demonstrated\cite{0957-4484-24-38-384004}.

Here, we use a theoretical model of reservoir computing called the {\em echo state network} (ESN) to show that reservoir computing can be employed to implement reliable and robust digital signal processing with  self-assembled nanowire networks. We extend the classical ESN with variable transfer functions and weight assignment to achieve a more realistic model of self-assembled networks. Compared to the RAC,  our system achieves 10 times higher success rate in implementing basic logic circuits and 2-bit adder and multiplier circuits.

%Reservoir computing is typically used for analog computation, in which the performance is  dependent on the dynamics of the reservoir, and is influenced by the reservoir structure and the statistics of the input signal, here the distributions of 1s and 0s in an input stream. We show that to implement combinational logic, where we only need the most recent inputs, the reservoir can be adjusted to be robust to the statistics of the input signal. We then present a detect-and-reconfigure algorithm that can detect component failure in the reservoir and reconfigure the output layer to regain  reliable computation.

%%%%%%%%
%%% Device
\section{Atomic Switch Networks}

{\em Atomic switch networks} (ASN) are a nanoscale technology based on silver nanowires developed by Terabe et al.\cite{terabe2005} aimed at reducing the cost and energy consumption of electronic devices. They can achieve a memory density of 2.5\,Gbit\,cm$^{-2}$ without any optimization, and  a switching frequency of 1\,GHz. Recently, Sillin et al.\cite{0957-4484-24-38-384004} combined bottom-up self-assembly and top-down patterning to self-assemble ASN. These networks are formed by electroless deposition of silver on pre-patterned copper seeds. Silver nanowires extended from the copper seeds form a three-dimensional structure that contains cross-bar-like junctions. When exposed to Sulfur gas, these junctions are transformed into metal-insulator-metal (MIM) interfaces and turned into atomic switches in the presence of external bias voltage. A silver filament forms to bridge the MIM interface, which increases the conductance across the switch, and is dissolved when the voltage is removed. The morphology of this self-assembled network can be directed by the pitch and the size of the copper seeds, which control the density and wire lengths, respectively. There are two types of variations among the switches: (1) the length of the MIM interface $w_0$, and (2) the growth rate of the filament in the presence of bias voltage $\alpha$. Both variations are modeled using a normal distribution with standard deviation up to $40\%$\cite{0957-4484-24-38-384004}. The I-V curve across each switch resembles that of a resistive switch, a saturated linear or semi-linear function. Self-assembled ASNs have been experimentally demonstrated for computing analog functions\cite{0957-4484-24-38-384004}. We use ASN as a model for physical RC implementation.

%%%%%%%%
%%% Models
\section{Models and Methods}

%%%%%%%
%%%%%%% Models: echo state network
\subsection{Reservoir Computing Model}
\label{sec:rc}

We use a well studied reservoir computing model called the {\em echo state network} (ESN)
\cite{Jaeger:2003p1447}. It consists of an input-driven recurrent neural network, which acts as the reservoir, and a readout layer that reads the reservoir states and produces the output. Unlike a classical recurrent neural network, where all the nodes are interconnected and their weights are determined during a training process, in an ESN the nodes are interconnected using random weights and random sparse connectivity between the nodes. The input and reservoir connections are initialized and fixed, with no further adaptation.

\begin{figure}[t]%{R}{0.5\textwidth}
\centering
\includegraphics[width=0.42\textwidth]{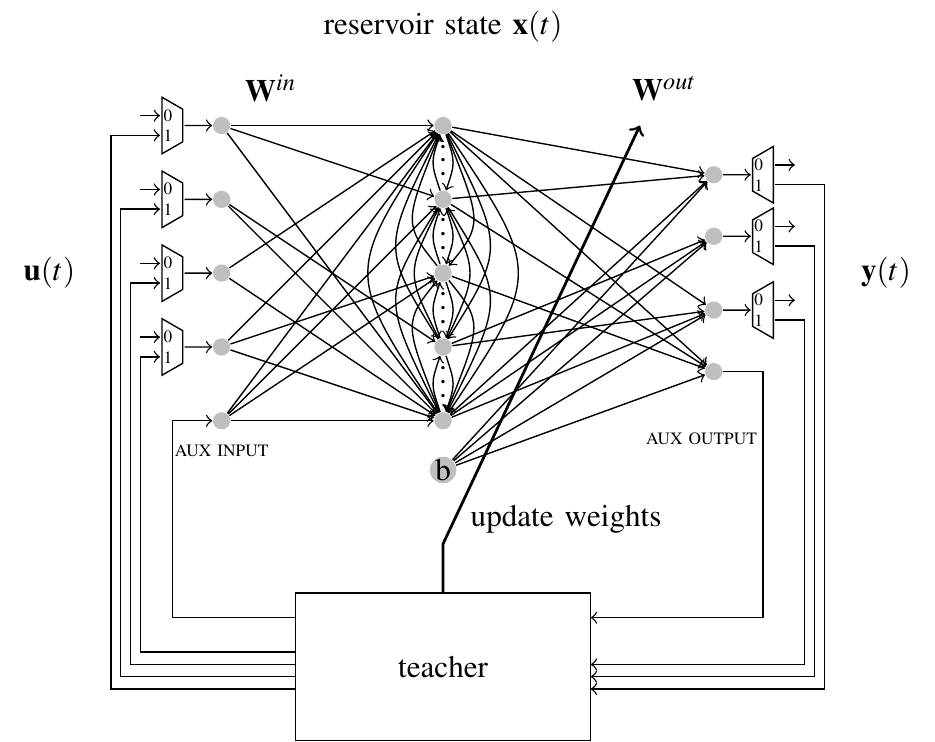}
\caption{Schematic of our augmented ESN. A dynamical core called a reservoir is driven by input signal ${\bf u}(t)$. The states of the reservoir ${\bf x}(t)$ are extended by a constant bias $b=1$ and combined linearly to produce the output ${\bf y}(t)$ To model self-assembled systems, the connectivity in the reservoir, between the inputs and the reservoir, and between the reservoir and the outputs is assigned randomly and sparsely. An auxiliary input and output is used to detect faults in the reservoir. The teacher unit contains stored data input-output pairs for all the input and output terminals. During the operation, the teacher feeds the auxiliary input with data and monitors the output with the matching target output. If the auxiliary output does not match the expected output, a failure has occurred in the reservoir. The teacher disconnects the reservoir from the external world and retrains the output weights, then reconnects the inputs and outputs to external signals.}
\label{fig:ESN}
\end{figure}

Figure~\ref{fig:ESN} shows a schematic of an ESN. The readout layer computes a linear combination of  reservoir states. The readout weights are determined using supervised learning techniques, where the network is driven by a teacher input and its output is compared with a corresponding teacher output to estimate the error; the weights can be calculated using any closed-form regression technique\cite{Jaeger:2003p1447} to minimize this error. We augmented the ESN with a teacher unit and auxiliary input and output to detect defects and retrain the network (see Section~\ref{sec:algorithm}). Mathematically, the input-driven reservoir is defined as follows. Let $N$ be the size of the reservoir. We represent the time-dependent inputs as a column vector ${\bf u}(t)$, the reservoir state as a column vector ${\bf x}(t)$, and the output as a column vector ${\bf y}(t)$. Rodan and Ti{\v n}o\cite{5629375} showed that the reservoir weight pattern does not affect the performance of the ESN. Here, we modify the classical ESN to model an ASN more closely. The input connectivity is represented by the matrix ${\bf W}^{in}$ where each element is assigned the weight $v$ with signs chosen according to Bernoulli distributions. A white noise with standard deviation 1 is added to the input weights to increase their variability. The reservoir connectivity is represented by an $N\times N$ weight matrix ${\bf W}^{res}$. A fraction $\delta_R$ of all possible connections are chosen to have weight one and the rest of the connections are assigned zero. The trade-off between nonlinearity and memory capacity of the reservoir is determined by the dynamics of the reservoir and can be adjusted by scaling the reservoir weight matrix as ${\bf W}^{res}\leftarrow\lambda{\bf W}^{res}/\lambda^{max}$, where $\lambda^{max}$ is the spectral radius of  ${\bf W}^{res}$ and $\lambda$ is the desired spectral radius $0<\lambda<1$; this range is necessary for the reservoir to adhere to the so-called echo state property, a condition that ensures that the long-term reservoir dynamics depends on the input signal and not on the initial state of the system.
The time evolution of the reservoir is given by:
\begin{equation}
{\bf x}(t+1) = f\left({\bf W}^{res}\cdot  {\bf x}(t) + {\bf W}^{in}\cdot {\bf u}(t)\right),
\end{equation}
where $f$ is the transfer function of the reservoir nodes that is applied element-wise to its operand. The output is generated by the multiplication of  an output weight matrix ${\bf W}^{out}$ of length  $N+1$ and the reservoir state vector $x(t)$ extended by a constant $1$ represented by ${\bf x}'(t)$:
\begin{equation}
{\bf y}_\tau(t) = {\bf W}^{out}\cdot {\bf x}'(t-\tau).
\label{eq:output}
\end{equation}

%%%%%%%
\noindent Here $\tau$ is output delay, the number of time steps to wait for the expected output to be ready. For brevity we omit writing $\tau$ in the subscript and explicitly mention it only when necessary. For training, we calculate the output weights to
minimize the squared output error $E=\langle ||{\bf y}(t)-{\bf \widehat{y}}(t)||^2 \rangle$ given the target output
${\bf \widehat{y}}(t)$. Here, $||\cdot||$ is the $L_2$ norm and $\langle\cdot\rangle$  the time average. The output weights are calculated using ordinary
linear regression as:

%%%%%%
%
\begin{equation}
{\bf W}^{out} = \left({\bf X}'{\bf X}-\gamma^2{\bf I}\right)^{-1}{\bf X}'{\bf Y},
\label{eq:regeq}
\end{equation}
where each row $t$ in the matrix ${\bf X}$ corresponds to the augmented state vector ${\bf x'}(t)$, $\gamma$ is a regularization factor, ${\bf I}$ is the identity matrix of order $N+1$, and ${\bf \widehat{Y}}$ is the target output matrix, whose rows correspond to target output vectors ${\bf \widehat{y}}(t)$. We use I, U, and N to refer to identical, uniformly distributed, and normally distributed reservoir weights, and we use $T$ and $L$ to refer to $\tanh$ or saturated linear transfer function given by:
\begin{equation}
f(x) = \begin{cases} 1, & \mbox{if } x\ge 1,\\
			     x, & \mbox{if } -1<x<1,\\
			     -1, & \mbox{if } x\le -1.
			     \end{cases}
\end{equation}
To study the sensitivity of the results to variations to the transfer functions, for each node we multiply the operand by a random variable with uniform distribution in the range $[0,2]$. We label such transfer functions with TV and LV for variable $\tanh$ or variable saturated linear function respectively. For example, a system with normally distributed   reservoir weights and variable saturated linear function is denoted by \esn{N}{LV}.

\subsection{Implementing a Universal Basis for Binary Logic}
We implement digital logic using reservoir computing by training the system as described above. Later, during the testing, we will threshold the output values at $\theta=0.5$ to obtain a digital signal.
To demonstrate the viability of reservoir computing for implementing general-purpose digital logic, we must show that any circuit can be implemented using reservoir computing components. To apply reservoir computing we have to pose our computations as a learning problem and in this setting some learning problems become more difficult to solve than in conventional top-down designed systems. An example is XOR, which is a linearly inseparable problem and cannot be learned easily. Yet, implementing XOR is necessary for many logic building blocks, such as parity-check and full-adder. One solution would be to implement a universal, but linearly separable logic function. Such a design raises two issues:
\begin{enumerate}
\item How many inputs can a reservoir of a given size process?
\item Can  reservoir computing be used for non-stationary signals?
\end{enumerate}
The performance of reservoir computing depends on the dynamics of the reservoir, which is influenced by the statistical properties of the input signal, such as its mean and standard deviation. Our intuition tells us that increasing the number of input signals will saturate the dynamics of the system to a point that it cannot be used by the output layer to construct the desired output. Moreover, we cannot guarantee that the signal statistics will remain constant over time during operation. To avoid this problem we use highly convergent reservoirs with low $\lambda$ so the only significant influence on the dynamics is from the most recent input. Through extensive simulation, we found that indeed in this regime the implemented logic is robust to changes in the input statistics. We can therefore avoid iterative algorithms and use  regular training. Additionally, this helps us to implement more complex logic from simpler ones, such as NAND, using a single reservoir by feeding back the output of the reservoir to itself. Ordinarily, the resulting system would require complicated training algorithms; however, insensitivity to signal statistics lets us treat the feedback connections as extra inputs. Currently, we do not have any analytical method to study the maximum number of inputs that can be processed with a reservoir of size $N$.  We will show some computational results regarding this question in Section~\ref{sec:result}.

\subsection{Variation and Fault Models}
\label{sec:variation}
Self-assembled nanoscale systems are known to be sensitive to environmental factors. They
affect  the structure of a nanoscale system in two ways: temporal and spatial variations in component properties and faults\cite{4447311}. To use reservoir computing for implementing logic functions, both problems must be addressed. To model  temporal variation we add a white noise term, with standard deviation $\sigma$, to $n$ randomly chosen non-zero entries of ${\bf W}^{res}$ at each time step $t$. Noise is added to the initial reservoir weight matrix at time $t$ and the noisy weight matrix is used to calculate the reservoir states for time $t+1$. This models variations in the electrical properties of nanowire networks   due to radiation or thermal noise. The normal distribution is known to be suitable to model variations in nanoscale devices\cite{4447311}. To study the effect of permanent failure, we randomly pick $m$ nodes and disconnect them from the network by setting the weights for all of their incoming and outgoing connections to zero. A similar approach was used in \cite{1049647,Lawson:2006-04-01T00:00:00:1546-1955:272}. In Section~\ref{sec:algorithm} we will show how such a permanent effect can be detected and treated.

\subsection{Design Space and Parameter Optimization}

For ESNs to work properly, there are numerous parameters that need to be optimized. Furthermore, our particular ESN model introduces more parameters to model self-assembled systems. This makes exhaustive exploration of the design space computationally prohibitive. For the purpose of demonstration, we used many preliminary experiments to develop heuristics and fix some of the parameters. We trained ESNs to compute 2-input to 10-input AND, NAND, OR, NOR, XOR, and XNOR gates. We are mostly interested in the implementation of NAND gates because of their universality, and the implementation of XOR gates for the complexity. We found that a reservoir size of $N=100$ nodes can compute 5-input NAND and XOR gates reliably. We use a training sequence of length $T=1000$ to train and to test ESNs. The choice of the parameter $\gamma$ depends on the statistics of ${\bf X'}{\bf X}$. The regularization factor $\gamma=0.015$ gives the best result across different tasks and network sizes, which we used for the rest of the study. Throughout this paper, the training and testing sequences were generated by sampling a binomial distribution with success probability of $p=0.5$. We studied the learning probability with respect to: (1) $\tanh$ and saturated linear transfer functions, with or without variation; (2) identical, uniformly distributed, and normally distributed reservoir weight patterns; (3) input coefficient $0.1<v<1.0$ and spectral radius $0.1<\lambda<1.0$; and, (4) input sparsity $\delta_I$, reservoir sparsity $\delta_R$, and output sparsity $\delta_O$ in the range $[0.1,1]$. We found that the optimal input coefficient and spectral radius across all the systems is $v=1.0$ and $\lambda=0.1$, regardless of the transfer functions, reservoir weight patterns, and sparsity of inputs, outputs, and the reservoir.  Unless explicitly mentioned, we used $n=1$ and $\sigma=0.1$ to induce temporal variations in the  networks.

\subsection{Reliability and Yield}
Self-assembly of nano-scale systems is an inherently stochastic process. Systems built using exactly the same process and parameters may differ structurally. Each fabricated system is a sample from a class of systems defined by the process parameters. Analogously, each instantiation of our model of RC, echo state networks, is a sample from a class defined by the combination of the parameters used. To argue that RC is a viable choice for building logic components using nanoscale systems, we have to show that ESN can implement digital logic reliably, i.e., with $100\%$ accuracy, and that  it can implement digital logic reliably with a high yield, i.e., the probability that a given instance  can achieve $100\%$ accuracy must be high. We can quantify this using the learning probability $LP$, the probability of perfect generalization given perfect training\cite{patarnello87:europhys}. We first calculate training and generalization accuracy $tr=\frac{1}{T}\sum_{t=1}^T \left(y_{tr}(t)-\widehat{y}_{tr}(t)\right)^2$, $gr=\frac{1}{T}\sum_{t=1}^T \left(y_{gr}(t)-\widehat{y}_{gr}(t)\right)^2$, where $T$ is the length of the input signal that drives the reservoir. We then define training and generalization probabilities $TP$ and $GP$ as the probability of perfect training and generalization accuracies:
\begin{align}
TP&=P[tr=1]=\frac{1}{\mathcal{N}}\sum_{i=1}^\mathcal{N} \lfloor tr_i \rfloor \text{ and}\\
GP&=P[gr=1]=\frac{1}{\mathcal{N}}\sum_{i=1}^\mathcal{N} \lfloor gr_i \rfloor.
\end{align}
Here $\mathcal{N}$ is the number of trials. The learning probability is given by $LP=P[gr=1|tr=1]=TP\times GP$. This is  justified because the input streams used in training and in testing are generated from independent distributions.

\subsection{Fault Detection and Retraining}
\label{sec:algorithm}
Self-assembled nanoscale systems are known to be susceptible to faults. Typically, a discovery and reconfiguration technique is used to find defective parts and to route communication around them to connect to all the functioning components. In the case of self-assembled nanowires, the components are simple switches and therefore reconfiguration techniques cannot be used. In RC, we only rely on the dynamics of the reservoir and adapt the output layer to interpret it properly. As long as the reservoir has dynamics, it can be used for signal processing. However, RC's  reliance on dynamics means that any change in the structure of the reservoir that affects the dynamical regime of the reservoir may have a catastrophic effect on the performance of the system. But since output weights can be calculated efficiently, we can design a teacher circuit that monitors the performance of the system and detects when the system becomes faulty. The teacher can then retrain the output layer. However, this means the teacher should have access, or at least should be able to compute, the correct output for any given input stream. But this begs the question. Our solution is to provide the reservoir with a pair of auxiliary input and output that can be used to detect a defective reservoir (Figure~\ref{fig:ESN}). The teacher can feed a stored input pattern through the auxiliary input signal and compare the auxiliary output with the corresponding stored target output. The auxiliary output is trained to produce the stored target output. A fault in the reservoir will cause a global disruption of the reservoir dynamics, almost certainly resulting in detectable erroneous output values for the auxiliary output. The teacher can then disconnect the main inputs and outputs from the external world and retrain the  output layer by feeding stored input streams for each input signal and recalculating the output weights using the matching stored target output according to Equation~\ref{eq:regeq}.

%%%%%%%%%
%%%%%%%%% Results
\section{Results}
\label{sec:result}
\begin{figure}[t]
\centering
\subfigure[\esn{I}{L}]{
\includegraphics[width=1in]{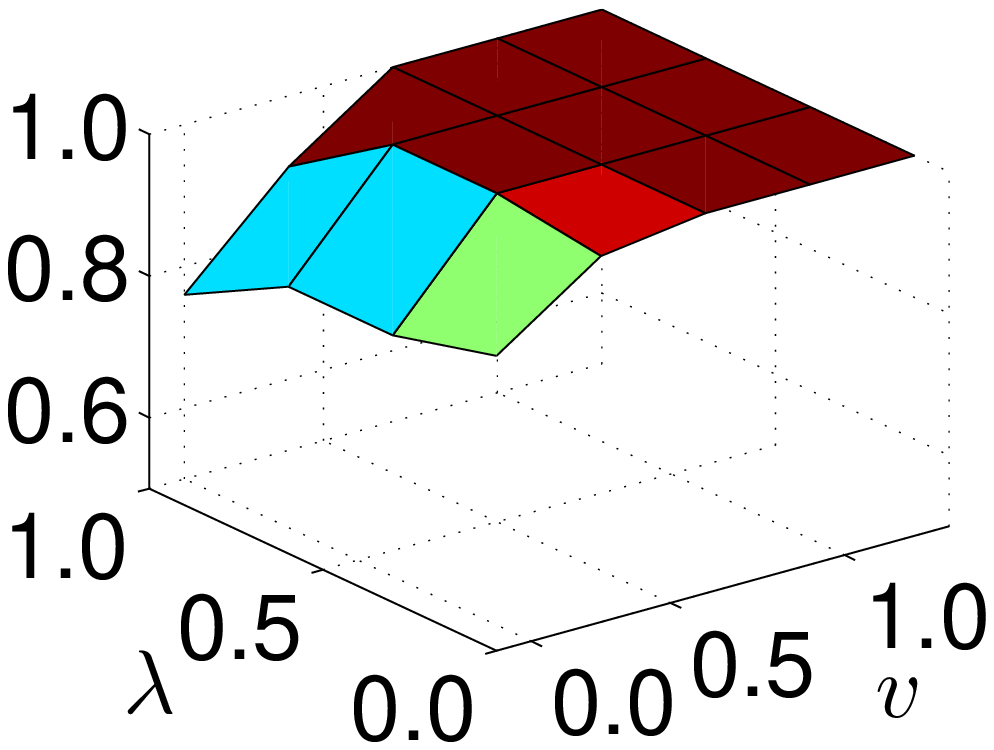}
}
\subfigure[\esn{U}{L}]{
\includegraphics[width=1in]{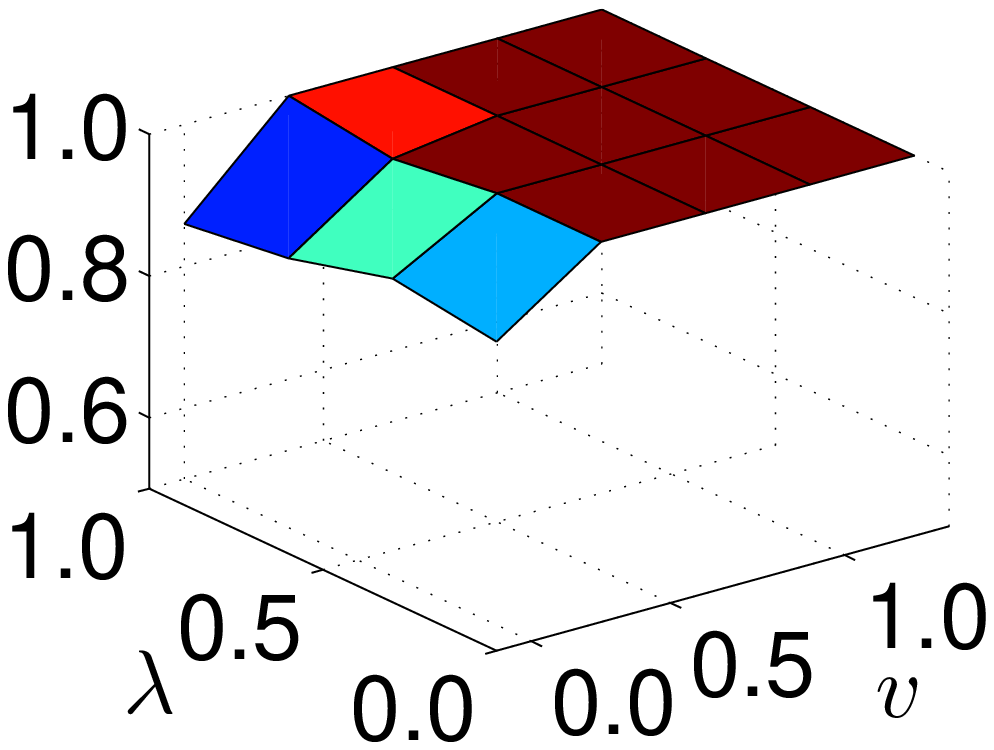}
}
\subfigure[\esn{N}{L}]{
\includegraphics[width=1in]{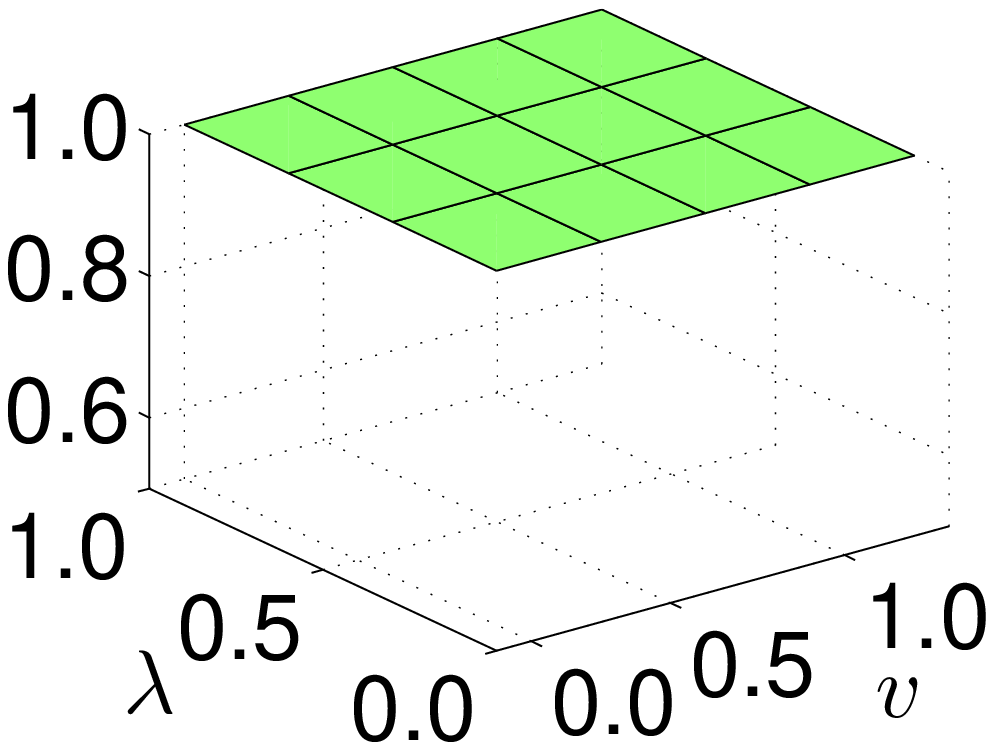}
}\\
\subfigure[\esn{I}{T}]{
\includegraphics[width=1in]{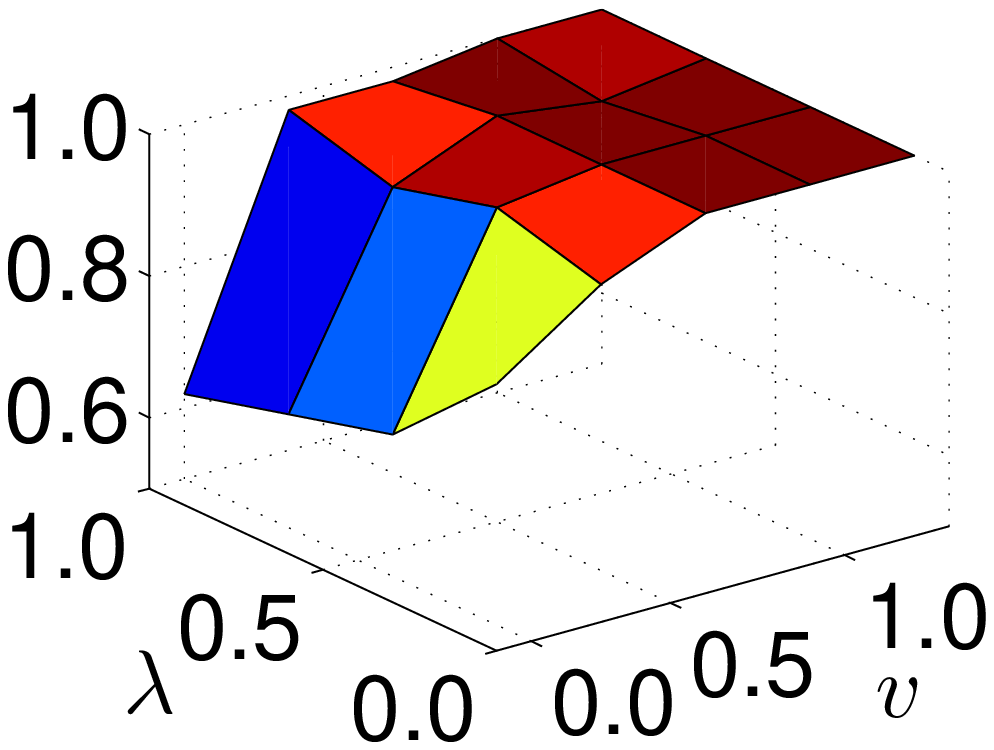}
}
\subfigure[\esn{U}{T}]{
\includegraphics[width=1in]{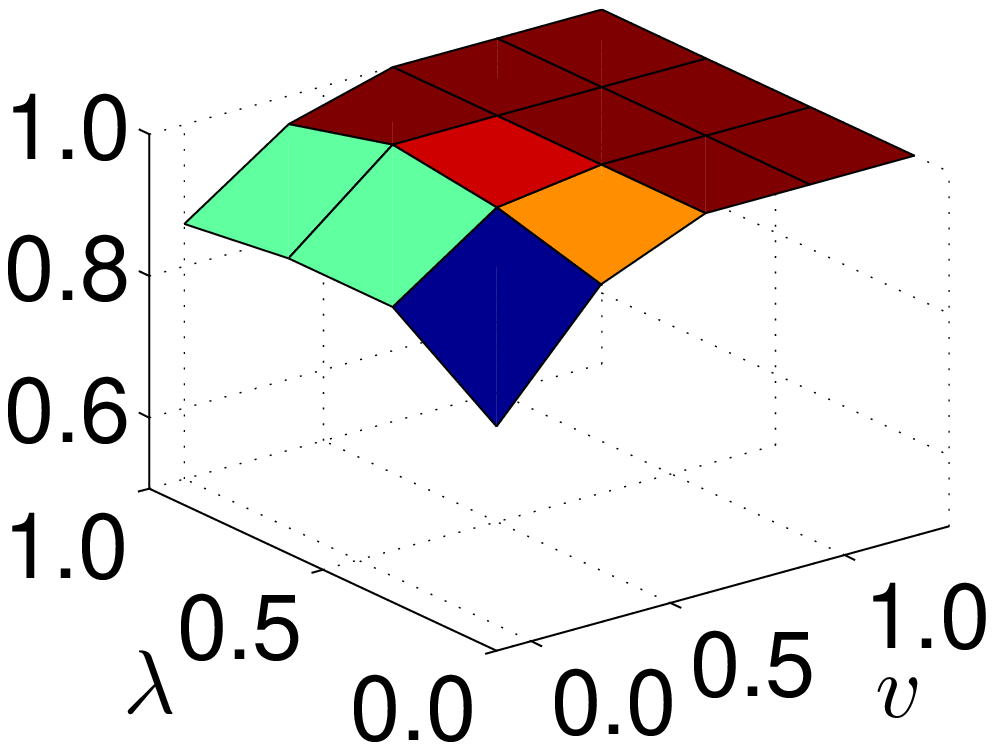}
}
\subfigure[\esn{N}{T}]{
\includegraphics[width=1in]{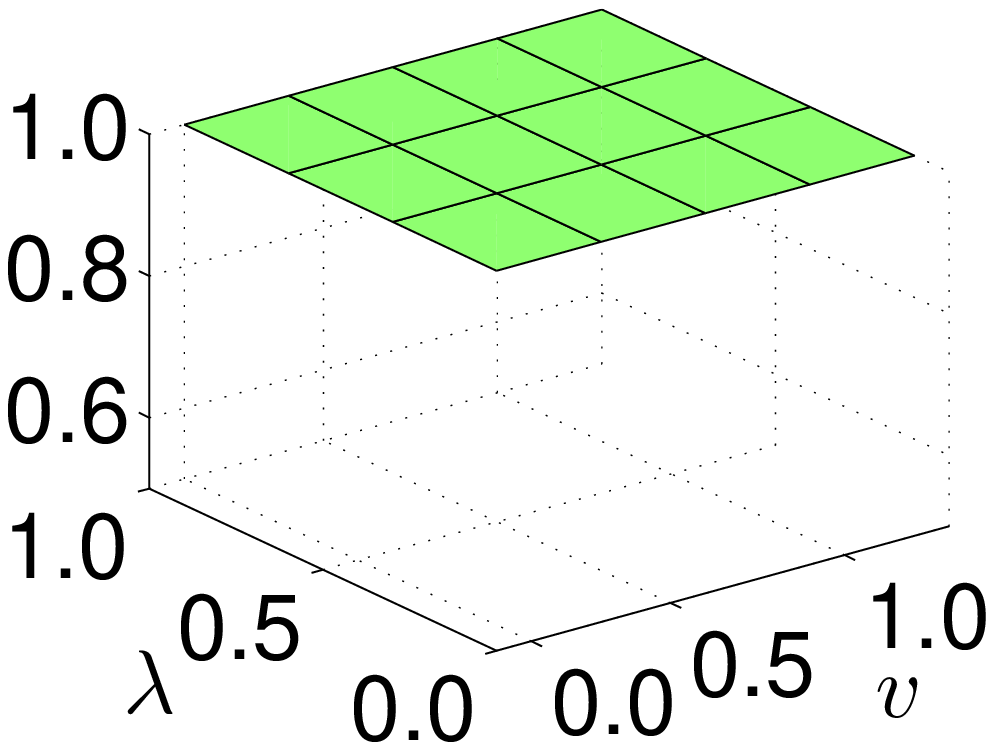}
}\\
\subfigure[\esn{I}{LV}]{
\includegraphics[width=1in]{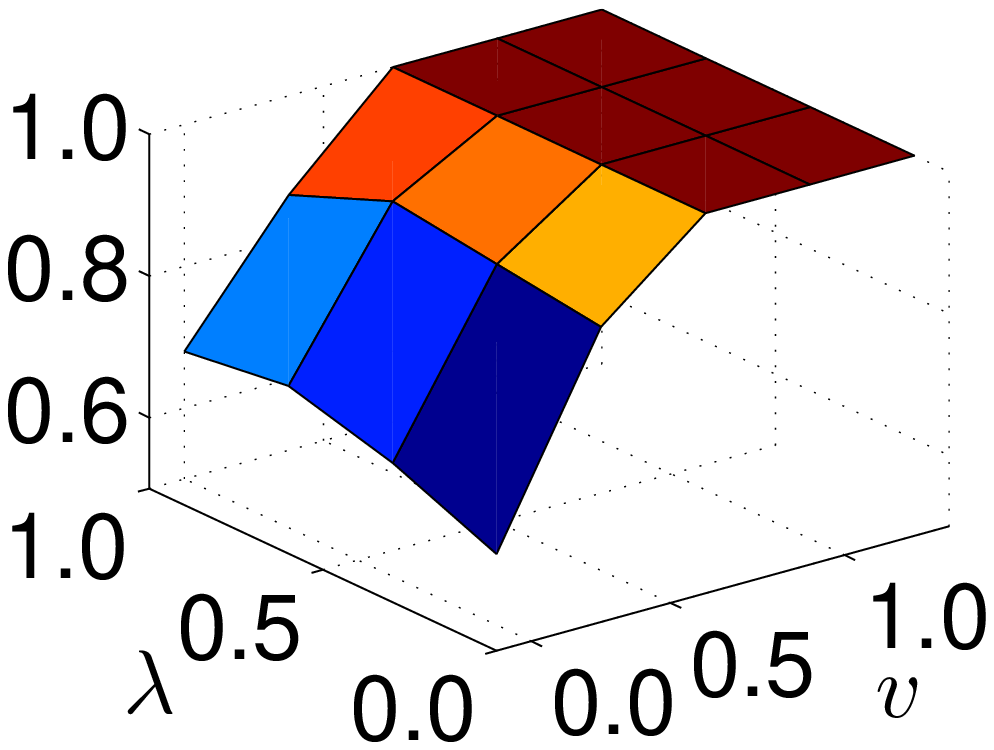}
}
\subfigure[\esn{U}{LV}]{
\includegraphics[width=1in]{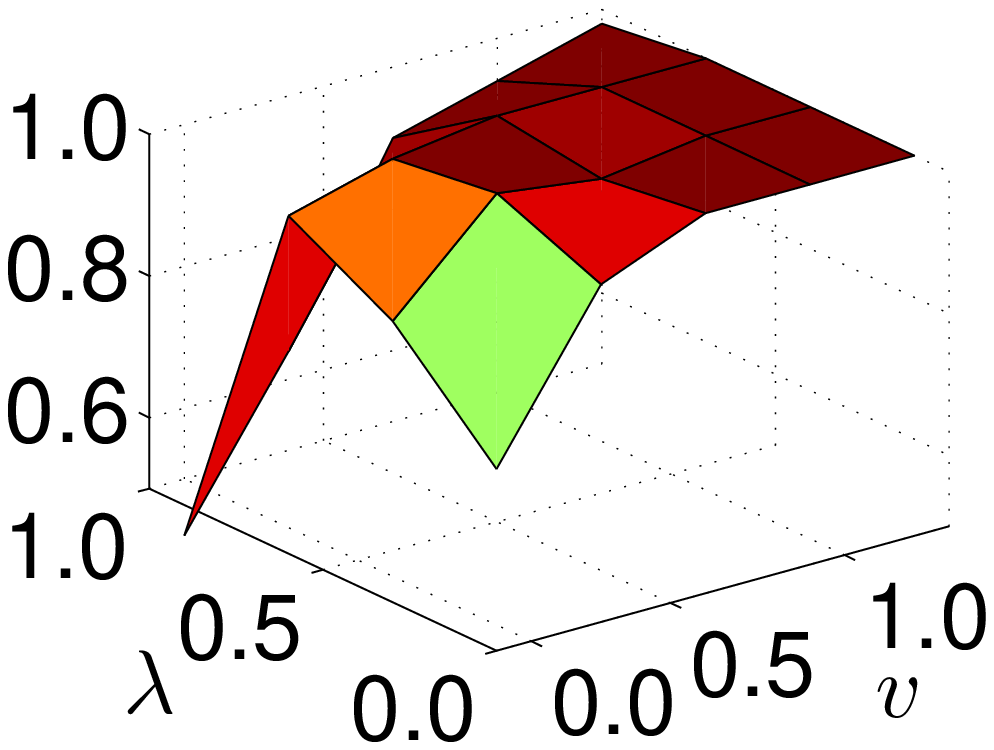}
}
\subfigure[\esn{N}{LV}]{
\includegraphics[width=1in]{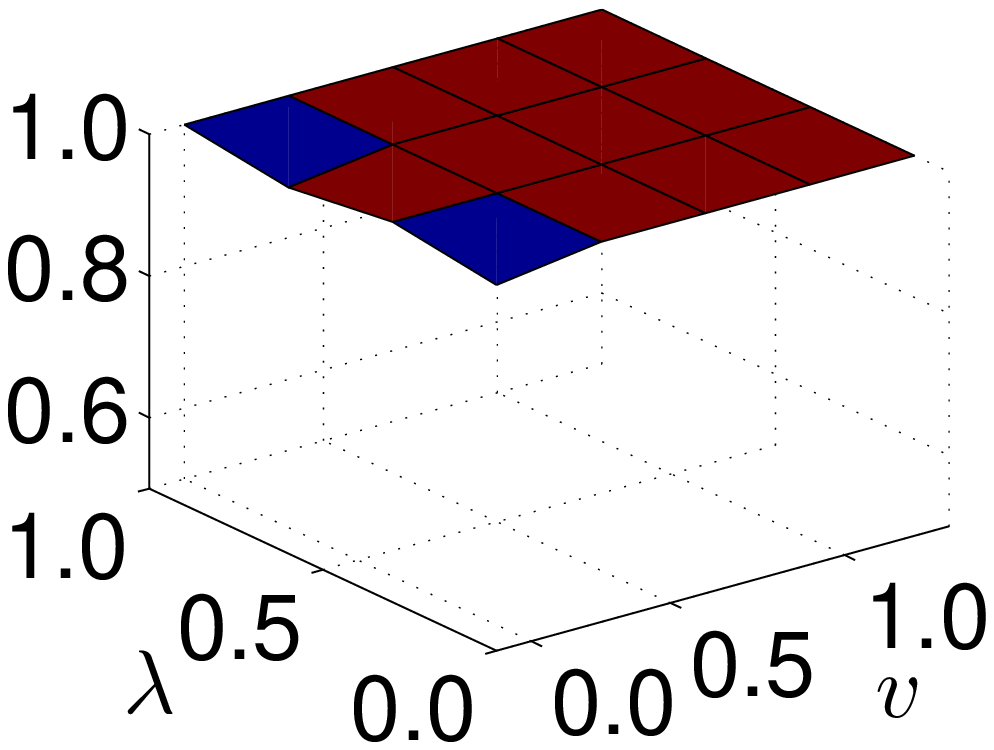}
}\\
\subfigure[\esn{I}{TV}]{
\includegraphics[width=1in]{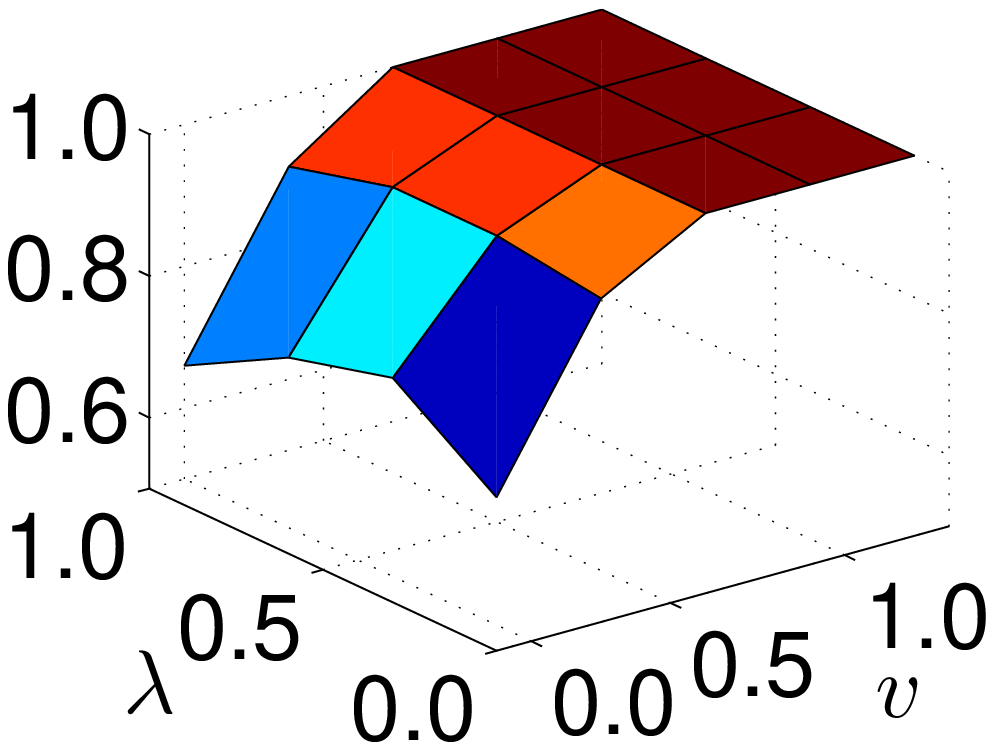}
}
\subfigure[\esn{U}{TV}]{
\includegraphics[width=1in]{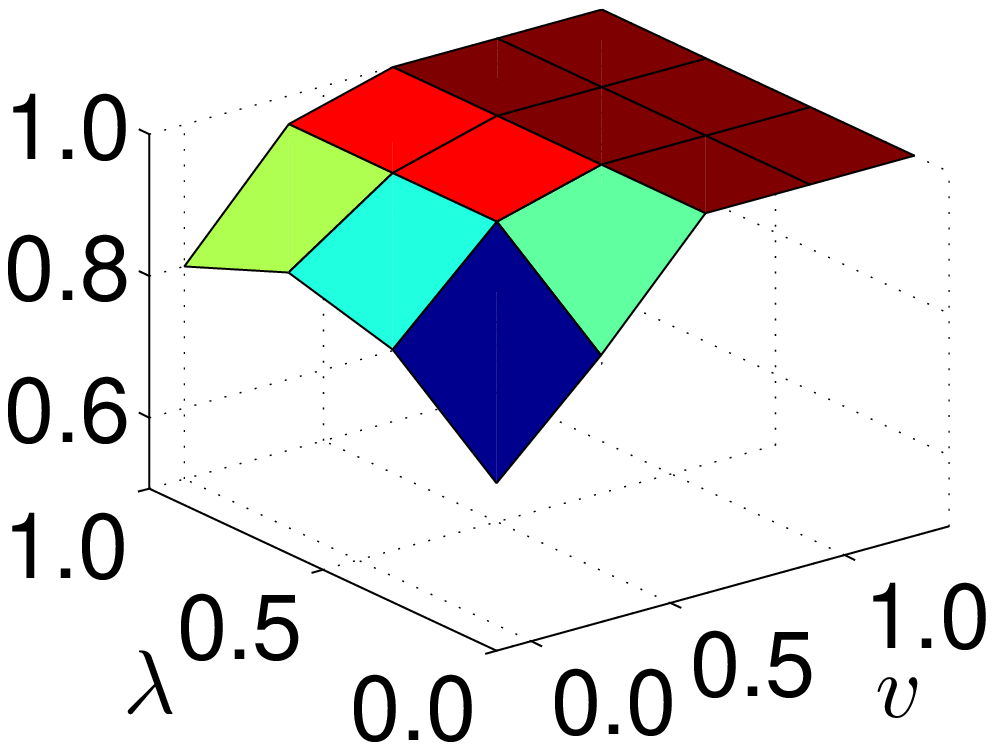}
}
\subfigure[\esn{N}{TV}]{
\includegraphics[width=1in]{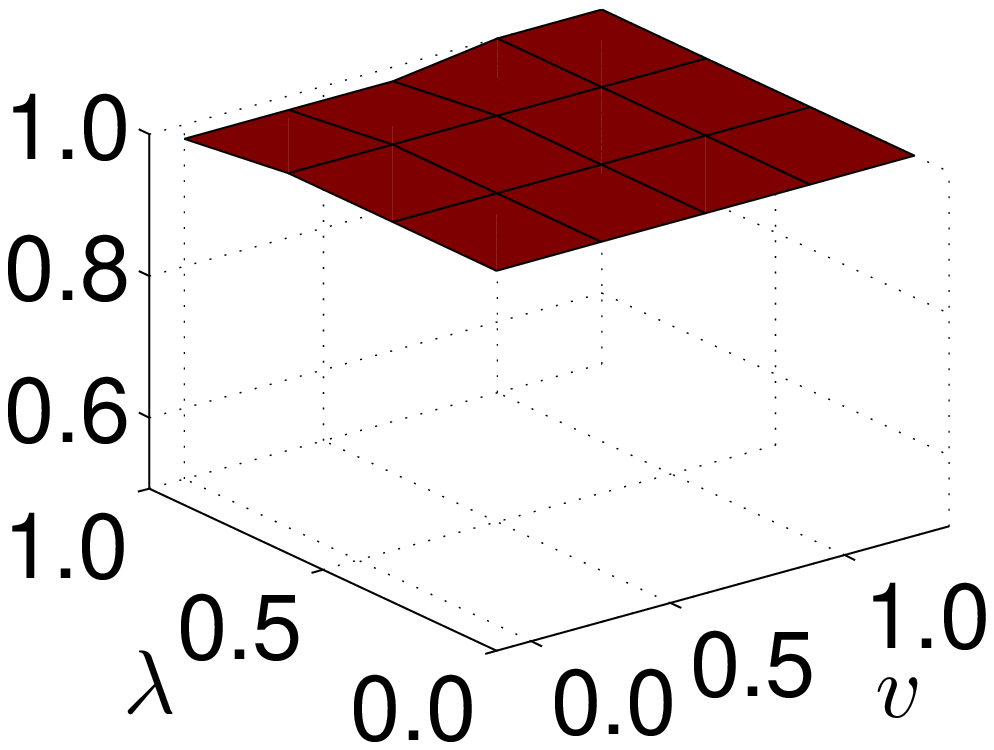}
}\\
\caption{Learning probability $LP$ of 5-input NAND as a function of $\lambda$ and $v$ for ESN with different
transfer functions and reservoir weights. Here I, U, and N refer to identical, uniformly distributed, and normally distributed weights, and L and T refer to saturated linear and $\tanh$ functions, and LV and TV refer to variable satured linear and variable $\tanh$ function. In all cases, optimal parameters are $v=1.0$ and $\lambda=0.1$. The same result hold for XOR, and for reservoirs with variable functions.}
\label{fig:lp1}
\end{figure}

We first demonstrate the insensitivity of optimal $v$ and $\lambda$ to reservoir weight patterns and transfer functions.
Unless otherwise mentioned, the results here are for $N=100$, $T=1,000$, and $\tau=1$. We specify the connectivity patterns when necessary. Figure~\ref{fig:lp1} shows the learning probability of 5-input NAND function for reservoirs with $\tanh$ and saturated linear functions and weight assignment of identical, uniformly distributed, and normally distributed. For this example, the reservoir is fully connected, and input and outputs are only connected to $75\%$ of the reservoir nodes. The figure also shows results of the experiments with variable saturated linear function (LV) and variable $\tanh$ functions (TV). We also performed the same experiment for 5-input XOR  and reservoirs with variable input, output, and reservoir sparsity. In all the cases the optimal parameters are $v=1.0$ and $\lambda=0.1$. For the rest of the experiments we use reservoirs with saturated linear functions and normally distributed weights, i.e., \esn{N}{L}.

%
%\begin{figure}[hb]
%\centering
%\subfigure[]{
%\includegraphics[width=1.6in]{nanoarch_tp_n300i4_nand}
%\label{fig:n300i4nandtp}
%}
%\subfigure[]{
%\includegraphics[width=1.6in]{nanoarch_tp_n300i5_nand}
%\label{fig:n300i5nandtp}
%}
%\caption{Training probability and learning probability as a function of $v$ and $\lambda$. Training probability of NAND gate for reservoirs of size 300 and 4 inputs (b) and 5 inputs (b).
%For 4 inputs, the $TP=1$ for a wide range of $v$ and $\lambda$; as the number of inputs increase to 5, the surface becomes more sensitive to changes in $v$ and $\lambda$ and peaks only in a narrow range.}
%\label{fig:tpn300}
%\end{figure}

\begin{figure}[b]
\centering
\subfigure[$\delta_R=1.0$]{
\includegraphics[width=1.6in]{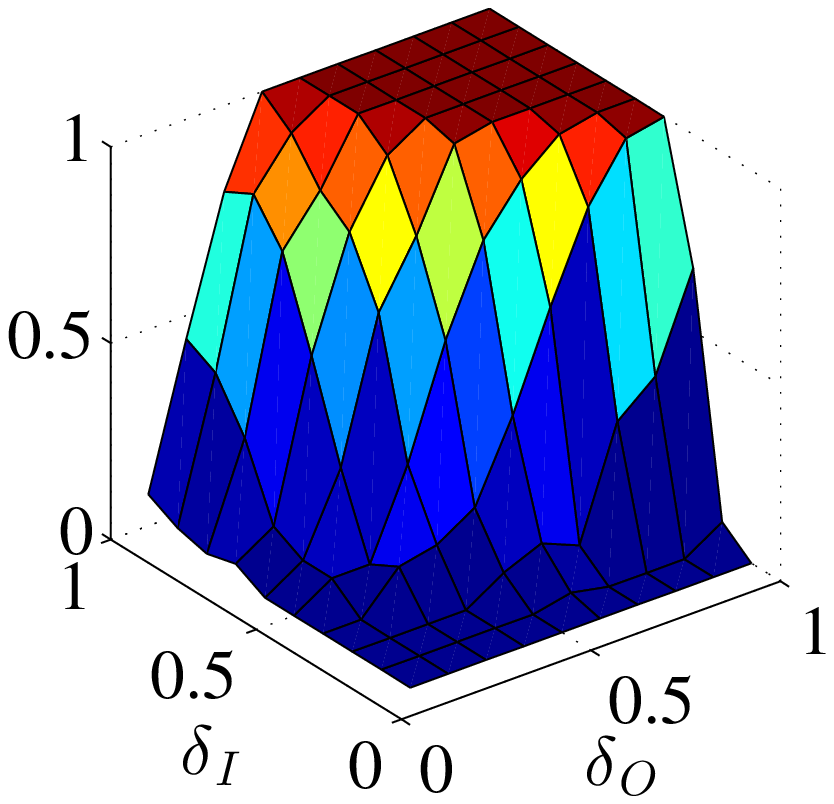}
\label{fig:insp}
}
\subfigure[$\delta_R=0.1$]{
\includegraphics[width=1.6in]{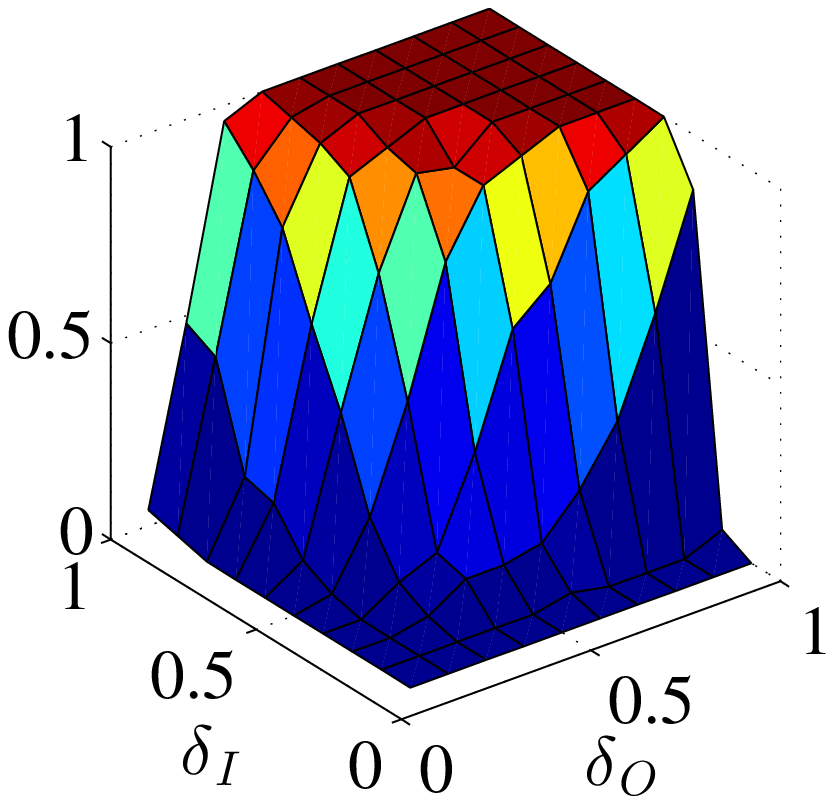}
\label{fig:rsp}
}
\caption{Learning probability $LP$ of 5-input NAND with optimal $v$ and $\lambda$ as a function of $\delta_I$ and $\delta_O$ for fully connected reservoirs (a) and reservoirs with connectivity fraction $\delta_R=0.1$ (b). The reservoir sparsity does not affect the $LP$.}
\label{fig:lp2}
\end{figure}

We then fix the $v$ and $\lambda$ to optimal values and study the effect of the input, output, and reservoir sparsity. Figure~\ref{fig:lp2} shows the $LP$ for 5-input  NAND as a function of $\delta_I$ and $\delta_O$ for fully connected reservoirs and reservoirs with connection fraction $\delta_R=0.1$. To achieve high learning probability $LP>0.5$, we need at least $\delta_I,\delta_O>0.4$. We use $\delta_I=0.5$ for the rest of the study. We also see that the reservoir connection fraction does not affect the $LP$.

\begin{figure}[t]
\centering
\subfigure[NAND]{
\includegraphics[width=1.6in]{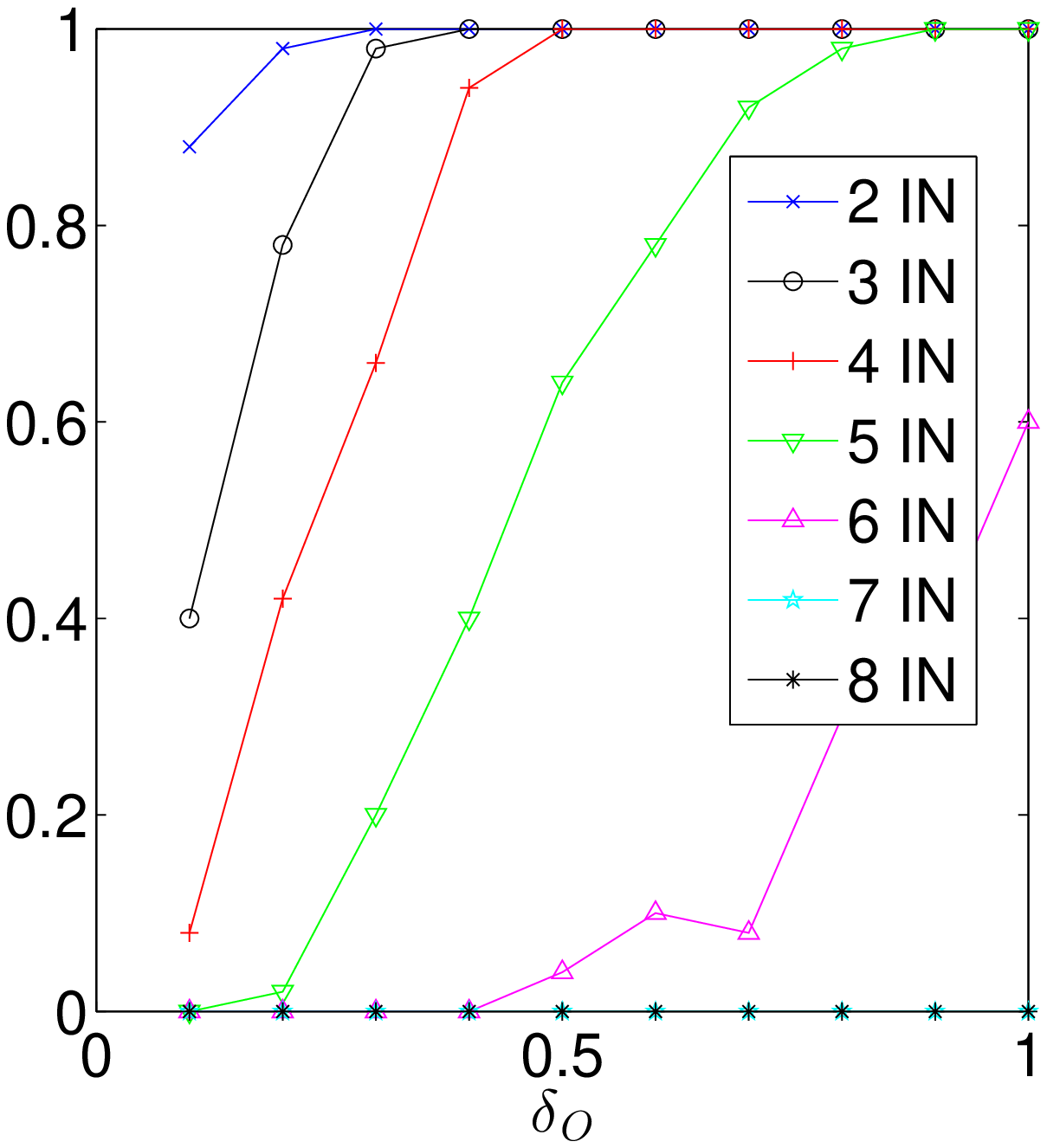}
\label{fig:nandvin}
}
\subfigure[XOR]{
\includegraphics[width=1.6in]{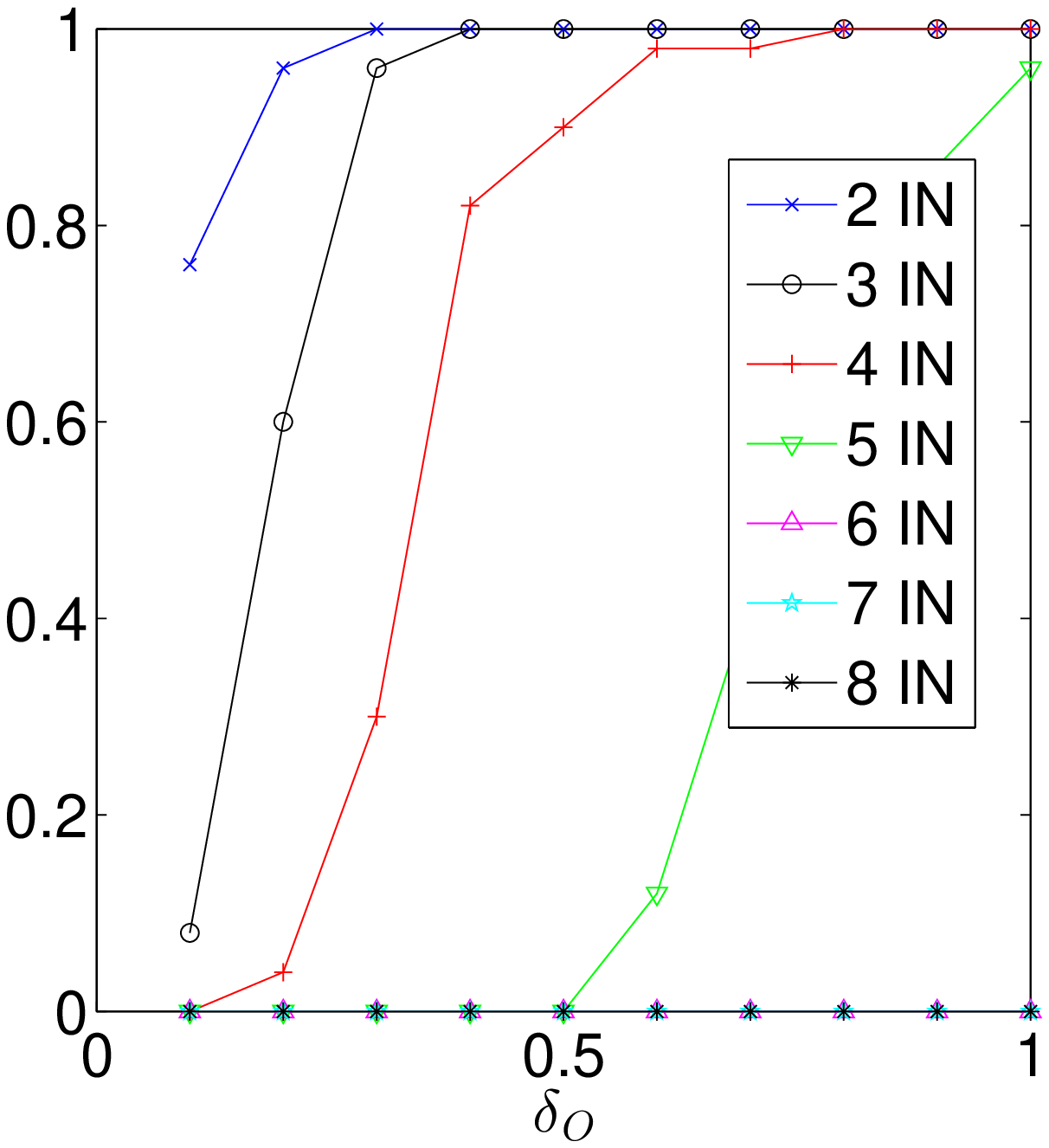}
\label{fig:xorvin}
}
\caption{Learning probability $LP$ of NAND as a function of  $\delta_O$ and number of inputs(a). Learning probability $LP$ of XOR as a function of  $\delta_O$ and number of inputs(b). In both cases we used reservoir size of $N=100$ with optimal $v$ and $\lambda$, $\delta_R=1$ and $\delta_I=0.5$.}
\label{fig:lp3}
\end{figure}

An important question  with a reservoir of fixed size: how does the performance of the system change as we increase the number of inputs? Figure~\ref{fig:lp3} shows the learning probability as a function of $\delta_O$ and the number of inputs. The task is to correctly implement a desired function, such as NAND and XOR, of all inputs. We observe that with a reservoir of $N=100$ nodes we can only compute NAND and XOR of up to 5 inputs with $LP=1$ and $LP=0.95$, respectively, even with dense output layer ($\delta_O=1$). Because of the nonlinearity of the XOR task, its performance degrades more quickly than NAND as the number of inputs increases. This can be compensated, to a degree, by the fact that we can produce all possible functions of 5 input at the same time. In fact, in all of our experiments we have used 6 outputs to compute OR, AND, XOR, and their negation in parallel. This demonstrates the ability of RC to compute multiple functions at the same time.

\begin{figure}[h]
\centering
\includegraphics[width=2.5in]{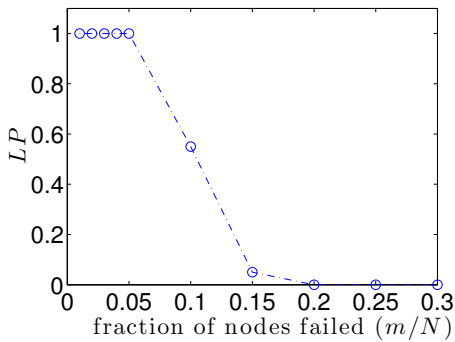}
\caption{Learning probability of main NAND after the failure recovery.}
\label{fig:failure}
\end{figure}

To compare our implementation to the RAC model, we use a \esn{N}{L} with $N=100$ nodes to implement a 2-bit adder and a 2-bit multiplier. In\cite{Lawson:2006-04-01T00:00:00:1546-1955:272}, networks of $N=100$ nodes and 40 control signals were used to implement these tasks; they achieved  the learning probability $LP=0.1$. In our experiments, we achieved $LP=0.6$, calculated over 500 trials, using output sparsity as low as $\delta_O=\delta_I=0.5$. Moreover, we computed both the adder and the multiplier in parallel using a single system; this uses half as many nodes compared with the RAC implementation.
Finally, to demonstrate the ability to detect and recover from permanent failure, we use \esn{N}{L} of $N=100$ nodes, and 4 inputs. We use 2 inputs as the main inputs and train the main output to compute their NAND. The other two inputs are used as auxiliary inputs. For simplicity, the auxiliary output is trained to compute the NAND of the two auxiliary inputs. We train the network with a stream of length $T=1,000$. During testing, at time step $t=700$ we pick $m$ nodes and disconnect them from the network as described in Section~\ref{sec:variation}. The teacher detects this failure and retrains the output layer. We repeat this experiment 20 times for each $m$ to study the effectiveness of our detect-and-retrain strategy. Figure~\ref{fig:failure} shows the result of this experiment. We measure the learning probability of the main NAND after the system has been retrained. We also see that for all $\frac{m}{N}<0.05$  the system regains complete performance in all of the experiments. All failure instances were detected in our experiments. This is an indication such a strategy works. We suspect that one must be careful in designing the function for the auxiliary output; choosing a simplistic function may not be effective for detecting the failures in the network.

%%%%%%%%%
%%%%%%%%% Results
\section{Discussion}

In this paper, we explored reservoir computing as an approach to computing with self-assembled nanoscale systems. Similar to Nanocell\cite{1049647} and RAC\cite{Lawson:2006-04-01T00:00:00:1546-1955:272}, the question we are addressing is: ``How can we use a disordered unreliable nanoscale system to compute reliably?" Compared with the Nanocell and the RAC, our approach does not require task dependent adaptation in the microstructure of the underlying system, nor does it require control signals to program the system. Because  programming does not affect the underlying system, each new output can be trained independently. This avoids the exponential
increase in training time as the number of outputs grows. Additionally, we are able to use the same system to implement multiple tasks in parallel. Our system achieves learning rates 6 times higher than the RAC, under variations five orders of magnitude higher than ITRS's projected target\cite{4447311}.

We demonstrated a retraining mechanism that can detect faults in the system and retrain the output layer to resume perfect operation. One limitation of this approach is that during  retraining the system is offline, therefore, system level fault-tolerance must be achieved using redundancy. Additionally, to achieve robustness to high variation, we must use a large system size. This can be compensated partially by training the system to have multiple outputs. The teacher circuit used for fault detection and retraining needs to have enough memory to store the input-output teacher data and to compute an inverse matrix, which is an $O(N^3)$ operation. Furthermore, we have assumed that the teacher can operate reliably.

%%%%%%%%%
%%%%%%%% Conclusion
\section{Conclusion}
We used a  model for reservoir computing to implement digital logic components and showed that the system can produce the desired logic with high probability, which translates to high yields in a fabrication process. However, our success comes at a price; compared with conventional top-down designed circuits, we used five times more switches to implement 5-input NAND logic. Our results indicates that reservoir computing is a viable approach for building digital systems that are tolerant to variations and faults using bottom-up self-assembled nanowires. This approach outperforms previously proposed models, and is more feasible since it does not require task specific modification to the system. We leave the complete characterization of the design space and device-level simulation using physically accurate models to future work.


\section*{Acknowledgment}
This material is based upon work supported by the National Science Foundation under grants 1028238 and 1028378. M.R.L. gratefully acknowledges support from the New Mexico Cancer Nanoscience and Microsystems Training Center (NIH/NCI grant 5R25CA153825).


\bibliographystyle{IEEEtran}
\bibliography{nanoarch}

\end{document}